\documentclass[onecolumn,nofootinbib,showpacs,aps,10pt]{revtex4}

\usepackage{amssymb}
\usepackage{amsmath}
\usepackage[dvips]{color}
\usepackage{bm}
\usepackage{graphics,graphicx}

\usepackage{xcolor}

\newcommand{\be}{\begin{equation}}
\newcommand{\ee}{\end{equation}}
\newcommand{\bea}{\begin{eqnarray}}
\newcommand{\eea}{\end{eqnarray}}

\newcommand{\slsh}[1]{\not \! #1}

\begin{document}

\title{Toward a Magnetic Warm Inflation Scenario}


\author{Gabriella Piccinelli$^{1,2}$\footnote{itzamna@unam.mx}, Angel S\'anchez$^2$\footnote{ansac@ciencias.unam.mx}}{\affiliation{$^1$Centro Tecnol\'ogico, Facultad de Estudios Superiores Arag\'on, Universidad Nacional Aut\'onoma de M\'exico, Avenida Rancho Seco S/N, Col. Impulsora Popular Av\'icola, Nezahualc\'oyotl, Estado de M\'exico 57130, M\'exico.
\\
$^2$Facultad de Ciencias, Universidad Nacional Aut\' onoma de M\' exico, Apartado Postal 50-542, Ciudad de M\'exico 04510, M\'exico.}

\begin{abstract}

In this work we explore the effects that a possible primordial magnetic field can have on the inflaton effective potential, taking as the underlying model a warm inflation scenario, based on global supersymmetry with a new-inflation-type potential. The decay scheme for the inflaton field is a two-step process of radiation production, where the inflaton couples to heavy intermediate superfields, which in turn interact with light particles. In this context, we consider that both sectors, heavy and light, are charged and work in the strong magnetic field approximation for the light fields. We find an analytical expression for the one-loop effective potential, for an arbitrary magnetic field strength, and show that the trend of the magnetic contribution is to make the potential flatter in the origin's vicinity, preserving the conditions for a successful inflationary process. This result is backed up by the behavior of slow-roll parameter $\epsilon$. The viability of this magnetic warm inflation scenario is also supported by the estimation of the effect of the magnetic field on the heavy particles decay width.

\end{abstract}


\maketitle

\section{Introduction}

{ The presently observed cosmic magnetic fields have an uncertain origin, that could be either astrophysical or primordial. The growing observational evidence of their existence in a wide range of spatial scales (see, e.g., \cite{Beck} and references therein) seems to indicate that they were probably present, at all times, during the universe evolution}. In this sense, it is important to take into account the magnetic field effect when addressing some early universe events. {  For example,  at the electroweak phase transition, the presence of an external hypermagnetic field modifies the phase transition making it stronger first order, lowering the critical temperature with respect to the non-magnetic case~\cite{ellos,autocitas}}. { In the case of the inflationary process, built on the grounds of an effective potential that describes a slow-roll transition, it has been shown that the effect of a magnetic field delays the transition~\cite{Warmus}: in the sense that the universe stays in the inflationary stage for longer time. In both cases, the magnetic field modifies the phase transition as in superconductors: as the magnetic field strength increases, the symmetry tends to be restored\footnote{ There is a subtle question regarding the dynamics of a slow-roll inflationary phase transition: The effective potential in particle physics is a quantity defined to describe static, equilibrium situations, that determines the ground state of the system, and hence the phase where it is located at. Nonetheless, it is used, in the standard picture of new inflation, as a description of an evolution: the slow-roll regime. Guth and Pi~\cite{GuthPi} analyzed this interpretation, tackling the problem with a different approach. They concluded that the na\"ive classical approach was essentially correct, within the parameters' domain of physical interest. Following this analysis, if the inflationary potential can be related with a system in equilibrium situation, the effect of the magnetic field on the system is to try to restore the symmetry, rising the secondary minimum as the magnetic field strength grows. In this sense, the situation is comparable to the superconductor phase, where the growth of the magnetic field intensity leads to the destruction of the superconducting phase.}. On another hand, the magnetic field also affects the particles decay process (see, e.g., \cite{JaberPiccinelliSanchez} and references therein).}

{ In an inflationary scenario,} if the magnetic field has to play any role, dissipative processes have to be considered. This could happen in models where the inflaton is coupled to gauge fields \cite{backs3}; to supersymmetric light fields, as in models of trapped inflation \cite{Green}; or to heavy ones, as in warm inflation \cite{Berera}. 
We are interested here in the warm inflation scenario, where the inflaton is assumed to interact with other fields during the whole inflationary process. Early models of inflation -dubbed super-cooled models (see e.g. \cite{Olive} for a review)- assumed very little interaction of the inflaton with all other fields until the reheating process, at the end of inflation. With the proposal of warm inflation \cite{Berera}, this picture changed: the inflaton is now assumed to interact with other fields, both during the inflationary expansion as well as at reheating, in a continuous and more natural way. It is a model where (near) thermal equilibrium conditions are maintained during the inflationary expansion, with no need for very flat potentials, nor for a tiny coupling constant. The model does require a dissipative component  of sizable strength as compared to the expansion rate of the universe. This is opposed to the standard inflationary scenario where the damping term comes only from the universe's expansion. A successful implementation of this model is embedded in the framework of supersymmetry, in order to ensure the cancellation of quantum fluctuations from fermions and bosons, protecting the flatness of the potential. Besides, it rests on a two-step process of radiation production, $\phi\rightarrow \chi \rightarrow \tilde{y}\tilde{y}$, where $\phi$ represents the inflaton, $\chi$ an intermediary heavy field and $\tilde{y}$ the light sector, composed of fermions $\psi_y$ and scalars $y$. In this way, the contribution from thermal corrections to the inflaton mass coming from heavy sector loops is Boltzmann suppressed. These $X$ fields are too heavy to be produced on shell and only appear as virtual $\chi$ (bosons) and $\psi_\chi$ (fermions) pairs, that decay into the light fields, that may or may not  thermalize \cite{HM, BereraRamos}. At the end, it is assumed that there is a soft SUSY breaking in the heavy sector. In this context, it has been shown that the quantum and radiative corrections do not spoil the slow-roll conditions required for inflation \cite{HM, Mar1} 

On the observational side, Planck's results establish a series of constraints on the abundant family of inflationary potentials proposed up to now \cite{Planck-Infl-2013, Planck-Infl-2018}. It is interesting to note that some potentials that are essentially ruled out by Planck in the context of cold inflation, are completely in agreement with observations when inflation evolves in a thermal bath \cite{being-warm, Enqvist}. Among single-field slow-roll inflationary models, warm inflation, although constrained \cite{Planck-NG-2015}, remains viable, the key feature being that the additional damping term introduced in this model can lower the tensor-to-scalar ratio $r$ for a given potential \cite{Planck-Infl-2018}. { However, this model faces with the requirement that a large number of degrees of freedom is needed to bring the system into a strong dissipative regime and still keep the amplitude of the primordial spectrum consistent with the observational value~\cite{microphysical}}.    

In a previous work \cite{Warmus}, we explored the effect of a weak magnetic field on the warm inflation effective potential, up to one loop, for neutral heavy bosons interacting with the charged light sector, showing that the magnetic field makes the potential flatter, retarding the transition{ : in the sense that the universe stays in the inflationary stage for longer time}, and works as an additional SUSY breaking scale. Here,  we broaden the scenario, allowing for magnetic fields of arbitrary strength and charged heavy fields. { Additionally, by estimating the magnetic field impact on the slow-roll parameters and the heavy particles decay width, we study the viability of this magnetic warm inflation scenario}. We locate our study in the context of warm inflation, in the sense that there is a light and a heavy sector from which energy is transferred. The thermalization of the light sector is not essential, as was pointed out in Ref.~\cite{BereraRamos}, but if it takes place, the temperature is in our case a lower scale as compared to the magnetic field, allowing to deal with a system at zero temperature. 

The paper is organized as follows: In Sec.\ref{sec2} we present a supersymmetric model we work with, in which all particles, except the inflaton field, are charged. In Sec.\ref{sec3}, we calculate the analytical expression for the inflaton one-loop effective potential for an arbitrary magnetic field strength within the model described in the previous section. In order to account for the magnetic field effect on the heavy particle masses, in Sec.\ref{sec4}, we compute the one-loop heavy field self-energy, coming from the light fields,  in the strong field limit, and estimate its decay width. In Sec.\ref{sec5}, we show our results through plots of the effective potential and the slow-roll parameters, for different values of the magnetic field. Finally, in Sec.\ref{sec6}, we present our conclusions.

\section{Model}\label{sec2}

Let us start by considering the equation of motion of the inflaton field that accounts a dissipation term ($\Gamma_\phi$) that comes from the interaction with a thermal bath and the universe expansion 
\begin{equation}
\ddot\phi+(3H+\Gamma_\phi)\dot\phi+V_{T,\phi}=0,\label{wip}
\end{equation}
where $H$ is the Hubble parameter,  and $V_{T, \phi}$ is the derivative with respect to $\phi$ of the inflaton effective potential (usually taken as the finite temperature one-loop Coleman-Weinberg potential). Warm inflation requires $\Gamma_\phi>3H$.

Since in warm inflation radiation is produced during the whole epoch of inflation, light fields, associated to this radiation, must be present in the Lagrangian. In these models, the inflaton interacts all the time with other fields, but its direct interaction with the light fields brings up some inconsistencies. Since the inflaton has a large expectation value, fields that interact directly with it acquire large masses. This fact is inconsistent with the radiation-like nature of such fields. Alternatively, one could limit the value of the coupling between the inflation and the light fields. However this would require an extremely low upper bound for the coupling, which in practice would make the interaction negligible. In view of these observations, a heavy field is introduced. This field acts as a mediator for the inflaton decay into light fields. This mechanisms results in a two step process of radiation production, $\phi\rightarrow \chi\rightarrow \tilde{y}\tilde{y}$, where $\phi$ represents the inflaton, $\chi$ the intermediary field and $\tilde{y}$ the light sector, composed of fermions $\Phi_y$ and scalars $y$. Also, in order to keep the flatness of the potential, one can resort to work within the framework of supersymmetry, since with such scenario, quantum fluctuations from fermions and bosons cancel out, which is a welcome feature to avoid spoiling the slow-roll conditions that are necessary for the flatness of the potential.

We start by extending the supersymmetric model used in Ref. \cite{HM} to a model where all particles are charged, except the inflaton field. With this in mind, the superpotential is
\be
W= -g\Phi X^2 -hXY_1Y_2,
\label{superpotential}
\ee
where $\Phi$, $X$ and $Y_{1,2}$ are chiral superfields, and  the second and the latter represent the heavy and light sector, respectively. The last term in the superpotential accounts for the interaction between  light and heavy sectors. 

The scalar interaction terms are derived from the superpotential in Eq. (\ref{superpotential}) as
\be
\mathcal{L}_S = -|\partial_\Phi W|^2 - |\partial_X W|^2 - |\partial_{Y_1} W|^2- |\partial_{Y_2} W|^2.
\ee
Defining $\phi=\sqrt{2}\ {\mbox{Re}}(\varphi)$, the inflaton potential, up to one loop correction is
\be
V(\phi)=\frac{1}{2}g^2 M_s^2 \left[\phi^2 \ln\left(\frac{\phi^2}{\phi_0}\right)+\phi^2_0-\phi^2\right],
\ee
where $\phi_0$ is the vev of the inflaton field.

The Yukawa sector is added to represent the interaction between the scalar and the fermionic sector:
\be
\mathcal{L}_{\mbox{\tiny{Yukawa}}}=-\frac{1}{2}\frac{\partial^2 W}{\partial \phi_n\partial \phi_m}\bar \psi_n P_L \psi_m
-\frac{1}{2}\frac{\partial^2 W^*}{\partial \phi^*_n\partial \phi^*_m}\bar\psi_n P_R \psi_m,
\ee
where $\phi_m$ is a superfield and
\be
P_{L,R}=(1\mp\gamma_5)/2.
\ee
Since fermion-boson cancellation takes place thanks to SUSY, the quantum corrections to the inflaton potential are shown to be small. Furthermore, the contribution from thermal corrections to the inflaton mass coming from heavy sector loops is Boltzmann suppressed. These $X$ fields are too heavy to be produced on shell and only appear as virtual $\chi$ (bosons) and $\psi_\chi$ (fermions) pairs, that decay into the light fields. We neglet the  heavy fields decay sub-leading process through $\chi \rightarrow y_1 y_2 \phi$ compared to $\chi \rightarrow \tilde y_1 \tilde y_2$, \cite{being-warm}.  It is also assumed that there is a soft SUSY breaking in the heavy sector and that light radiation thermalizes.

The set of interactions that involve the inflaton and the $\chi$ field can be obtained from the scalar (${\cal L}_s$) and fermion (${\cal L}_f$) sectors of the Lagrangian, given by
\begin{eqnarray} \nonumber
{\cal L}_s  &=& g^2 |\chi|^4 +4g^2 |\varphi|^2|\chi|^2\\
&+&h^2(|y_1|^2|+|y_1|^2)|\chi|^2+h^2|y_1|^2|y_2|^2
+2gh(y_1y_2\varphi^\dagger\chi^\dagger+y_1^{\dagger}y_2^{\dagger}\varphi\chi)
\nonumber \\   
{\cal L}_f &= &g(\varphi \overline \psi_{\chi} P_L\psi_{\chi}
+\varphi^\dagger \overline \psi_{\chi}P_R \psi_{\chi})
+2g(\chi\overline\psi_{\chi}P_L \psi_{\varphi} +
\chi^\dagger\overline\psi_{\chi} P_R \psi_{\varphi}) 
\nonumber \\ 
&+&h\chi(\overline\psi_{y_1}P_L\psi_{y_2}+\overline\psi_{y_2}P_L\psi_{y_1}) + h\chi^\dagger( \overline\psi_{y_1} P_R \psi_{y_2}+\overline\psi_{y_2} P_R \psi_{y_1}) \nonumber \\ 
&+&2h(y_1\overline\psi_{y_2} P_L\psi_{\chi}+y_2\overline\psi_{y_1}P_L\psi_{\chi})  +2h(y_1^\dagger\overline\psi_{y_2}P_R\psi_{\chi}+y_2^\dagger\overline\psi_{y_1}P_R\psi_{\chi}),
\label{lagrangian}
\end{eqnarray}
where $\varphi$, $\chi$ and $y_{1,2}$ are the scalar field component of the chiral  superfields $\Phi$, $X$ and $Y_{1,2}$, respectively. $\psi_i$ denotes the fermion fields coming from the different sectors and $g$ and $h$ are coupling constants, whose values, limited by the slow-roll conditions and the constraints from density perturbations, are $\mathcal{O}(0.1)$ \cite{HM}. 

The effect of a magnetic field on the inflationary process can be accounted for in two ways: through heavy fields effective potential and their quantum fluctuations due to the interaction with the light sector. With these ideas in mind, we shall calculate the magnetic contributions to the heavy fields one-loop potential in vacuum, since these fields are so heavy that their contribution from thermal corrections to the inflaton mass, coming from this sector loops, is Boltzmann suppressed. By doing this, we are considering that the magnetic field strength can be of the same order of magnitude as the heavy masses. On another hand, since the magnetic field is the highest scale, the thermal contributions on the heavy sector masses coming from the interaction with the light particles are also neglected.

\section{Effective potential}\label{sec3}

Taking into account that the heavy fields are fermionic and bosonic, the contribution to the inflaton effective potential coming from this sector, up to one loop, reads
\bea
 \mathcal{V}^1(\phi)=V^1_{\chi}+V^1_{\psi_{\chi}},
\eea
which, in absence of an external magnetic field, have the form   
\bea
    V^1_{\chi,0}&=&\frac{4}{2}\int \frac{d^4p}{(2\pi)^4} \ln (p^2-m_\chi^2),\\
    V^1_{\psi_\chi,0}&=&\frac{4}{2} \int \frac{d^4p}{(2\pi)^4} \ln(p^2-m_{\psi_\chi}^2),
\eea
with $m_\chi$ and $m_{\psi_\chi}$ the boson and fermion masses, respectively; fermions are Weyl spinors with two degrees of freedom, and the coefficients account for charge and spins. The subscript ``0" emphasizes the absence of the external magnetic field.

In the presence of an external and uniform magnetic field $B$, which defines the $z$-direction, the above expressions become
\bea
    V^1_{\chi}&=&\frac{4}{2}\int \frac{d^4p}{(2\pi)^4} \left(\ln D^{-1}_B(p)\right),\\
    V^1_{\psi_\chi}&=&\frac{1}{2} \int \frac{d^4p}{(2\pi)^4}\ln Det(S^{-1}_B(p)),
    \label{potfer1loop}
\eea 
where $D$ and $S$ are the propagators of boson and fermion, respectively, and are given by
\bea
\label{scalpropmom}
D_{B} (p) &=& \int_0^\infty \frac{ds}{\cos{q_\chi Bs}} \exp\left\{ i s (p_{||}^2-p_{\bot}^2 \frac{\tan{q_\chi Bs}}{q_\chi Bs}-m^2_\chi +i
\epsilon)\right\}, \\
S_{B} (p)&=&\int_0^\infty \frac{ds}{\cos{q_\chi Bs}} \exp\left\{ i s (p_{||}^2-p_{\bot}^2 \frac{\tan{q_\chi Bs}}{ q_\chi Bs}-m^2_{{\psi_\chi}}
+i \epsilon)\right\} \left[ (m_{\psi_\chi}+{\not \! p}_{||})e^{i q_\chi B s \Sigma_3} -\frac{{\not \!
p_\bot}}{\cos{q_\chi B s}}\right],
\label{ferpropmom}
\end{eqnarray}
with $s$ the Schwinger's proper time parameter, $(a\cdot b)_{||}\equiv a_0b_0-a_3b_3$, $(a\cdot b)_\perp\equiv a_1b_1+a_2b_2$  and $\Sigma_3\equiv i\gamma^1\gamma^2$. $q_\chi$ denotes the charge associated to the heavy superfield fermion or boson components. 

Once the integration over the momentum is carried out, and all divergent terms are isolated,  the effective potential  can be rewritten as
\bea
   \mathcal{V}^1(\phi,B)&=& \mathcal{V}^1_{0}+\mathcal{V}^1_{q_\chi B^2}+\mathcal{V}^1_{df},
   \label{potentialfull}
 \eea
 with 
 \bea
    \mathcal{V}^1_{0}&=&\frac{1}{8\pi^2}\int_0^\infty \frac{ds}{s^3}\left\{ e^{-s m_\chi^2} -e^{-s m_{\psi_\chi}^2}\right\},  
\eea
\bea
   \mathcal{V}^1_{q_\chi B^2}&=&-\frac{1}{8\pi^2}\int_0^\infty \frac{ds}{s}\left\{ e^{- s m_\chi^2} + 2\ e^{- s m_{\psi_\chi}^2} \right\}\frac{(q_\chi B)^2}{6}, 
\eea

\bea
   \mathcal{V}^1_{df}&=&\frac{1}{8 \pi^2} \int_0^\infty \frac{ds}{s^3}
   \left\{
      e^{-sm_\chi^2}
   \left[\frac{q_\chi B s}{\sinh(q_\chi B s)}-1+\frac{1}{6}(q_\chi Bs)^2\right]
  -e^{-sm_{\psi_\chi}^2}
   \left[q_\chi B s \coth (q_\chi B s)-1-\frac{1}{3}(q_\chi Bs)^2\right]
\right\},
\label{vfinite_s}
\eea
where the masses $m_\chi$ and $m_{\psi_\chi}$ keep track of the bosonic and fermionic sectors, respectively. Notice that the effective potential in Eq.~(\ref{vfinite_s}) is divergence free ($df$)~\cite{Landaucourse4}, meanwhile, the first two contributions to the effective potential are divergent when SUSY is broken. However, since we are considering the external magnetic field as a classical field, then the energy of the quantum fluctuations cannot go beyond a certain scale $\Lambda$. Thus, in these two integrals we have to introduce this ultraviolet cutoff.  By doing this, it is not difficult to show that each expression goes like
\bea
      \mathcal{V}^1_0&=&-\frac{1}{32\pi^2} \left\{m_\chi^4 \ln\left(\frac{m^2_\chi}{\Lambda^2}\right)- m_{\psi_\chi}^4 \ln\left(\frac{m^2_{\psi_\chi}}{\Lambda^2}\right)\right\}+C,
      \label{potV0}
\\      
        \mathcal{V}^1_{q_\chi B^2}&=&-\frac{1}{8\pi^2}\frac{(q_\chi B)^2}{6} \left\{\left[-\gamma_E - \ln\left(\frac{m^2_\chi}{\Lambda^2}\right)\right]+2\left[-\gamma_E -\ln\left(\frac{m^2_{\psi_\chi}}{\Lambda^2}\right)\right]\right\},
\eea
where $C$ is a constant which can be determined from the renormalization conditions.
Note that the main divergences cancel out and the remaining ones are due to the soft SUSY breaking term which we have defined as the slight difference between the fermion and boson masses, that is
\begin{eqnarray}
m^2_\chi(B)&=& 2g^2\phi^2+m_b^2(B) + M_s^2, \nonumber \\
m^2_{\Psi_\chi}(B)&=& 2g^2\phi^2 + m_f^2(B),
\label{eq:susybreaking}
\end{eqnarray}
where  $m_b^2(B)$ and $m_f^2(B)$ are the magnetic one-loop self-energy corrections to the fermion and boson  masses, respectively.

The integral over the proper time in Eq.(\ref{vfinite_s}) can be exactly done, obtaining
\bea
\mathcal{V}^1_{df}&=&\frac{(q_\chi B)^2}{8\pi^2}\left\{ 
   \left[4 \zeta ^{(1,0)}\left(-1,\frac{m_\chi^2}{2 {q_\chi B}}\right)-2 \zeta
   ^{(1,0)}\left(-1,\frac{m_\chi^2}{{q_\chi B}}\right)+\frac{m_\chi^4 \left(2 \ln
   \left(\frac{2m_\chi^2}{{q_\chi B}}\right)-1\right)}{4 (q_\chi B)^2}-\frac{ m_\chi^2 \ln(4)}{2 {q_\chi B}}-\frac{1}{6} \ln \left(\frac{m_\chi^2}{4
   {q_\chi B}}\right)-\frac{1}{6} \right]\right.\nonumber \\
   &&\ \ \ -\left.\left[-4 \zeta ^{(1,0)}\left(-1,\frac{{m_{\psi_\chi}}^2}{2 {q_\chi B}}\right)+\frac{{m_{\psi_\chi}}^4 \left(2 \ln
   \left(\frac{{m_{\psi_\chi}}^2}{2 {q_\chi B}}\right)-1\right)}{4 (q_\chi B)^2}-\frac{{m_{\psi_\chi}}^2 \ln
   \left(\frac{{m_{\psi_\chi}}^2}{2 {q_\chi B}}\right)}{{q_\chi B}}+\frac{1}{3} \ln \left(\frac{{m_{\psi_\chi}}^2}{2
   {q_\chi B}}\right)+\frac{1}{3}\right] \right\}.
 \nonumber \\
\eea
In order to determine the $\Lambda$ and $C$ values, we impose that at $B=0$, the effective potential lower value be zero at $\phi=\phi_0$, with $\phi_0$ the inflaton $vev$, that is 
\bea
     \left.\mathcal{V}^{1}(\phi,B)\right|_{\phi=\phi_0}=0
\eea 
and
\bea
     \left.\frac{\partial}{\partial \phi}\mathcal{V}^{1}(\phi,B)\right|_{\phi=\phi_0}=0,
\eea  
getting
\bea
   \ln\left(\frac{\Lambda^2}{2g^2\phi_0^2+M_s^2}\right)=\frac{1}{2}+\frac{2g^2\phi_0^2}{M_s^2}\ln\left( \frac{2g^2\phi^2_0+M_s^2}{2g^2\phi^2_0}\right)
\eea
and 
\bea
    C=\frac{g^2\phi_0^2 M_s^2}{32\pi^2}
    \left[\frac{4g^2\phi_0^2+M_s^2}{2g^2\phi_0^2}+2\frac{2g^2\phi_0^2+M_s^2}{M_s^2}\ln\left(\frac{2g^2\phi^2_0+M_s^2}{2g^2\phi^2_0}\right)\right].
\eea
In the next section we shall calculate the magnetic contribution to the heavy particle masses through the light quantum fluctuations.

\section{Heavy Particles self-energy}\label{sec4}

According to Eq.(\ref{lagrangian}), there are four Feynman diagrams in which the heavy fields interact with light particles. In what follows we calculate each one of them in the presence of a uniform magnetic field, bearing in mind that this external magnetic field is the highest physical scale compared with the light particle parameters, as temperature and masses.

\subsection{Magnetic Masses}
\subsubsection{Scalar self-energy dressed with magnetic field effects}

In the case of the heavy scalar field there are two different vertices that couple them with the light fields. Let us start with the simplest interaction, displayed in Fig.~\ref{FDfigures}a, representing the interaction between two heavy and two light scalar fields. In the configuration space, it has associated the following mathematical expression
\bea
   -i \Sigma(w,w) &=& (-i 4h^2 ) \frac{1}{2}\sum_{i} D_{y_i}^B(w, w),
\eea
where the sum over $i$ accounts for the two light boson species and the symmetry factor has been accounted for. 
In the momentum space, the above equation reads
\bea
    \Sigma(p) &=&  2h^2 \sum_{i}\int \frac{d^4k}{(2\pi)^4} D^B_{y_i}(k), 
\label{LLLtad1}
\eea
with $D^B_{y_i}(k)$ the scalar propagator given in Eq.~(\ref{scalpropmom}). This propagator can be rewritten in terms of the Landau levels by doing a deformation in the integration contour over the proper time $s$, as shown in~\cite{Ayala2005}.  

\begin{figure}
\begin{tabular}{ccc}
     \includegraphics[width=4cm]{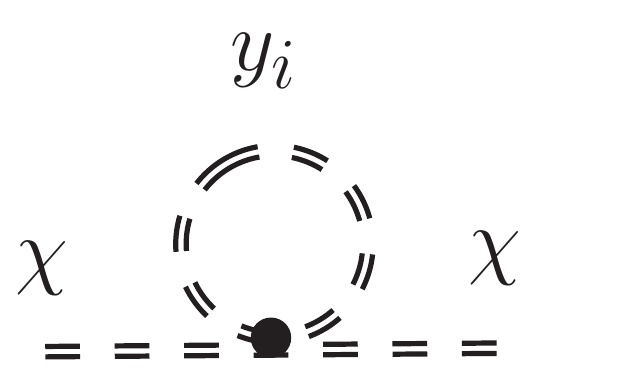} & \hspace{2em} & \includegraphics[width=4cm]{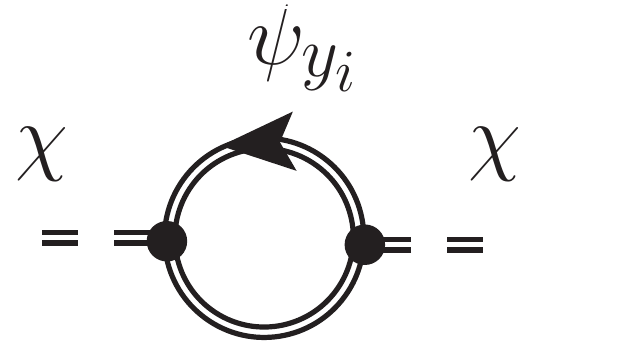} \\
     (a) & \hspace{2em} & (b) \\
     & &  \\
     \includegraphics[width=4cm]{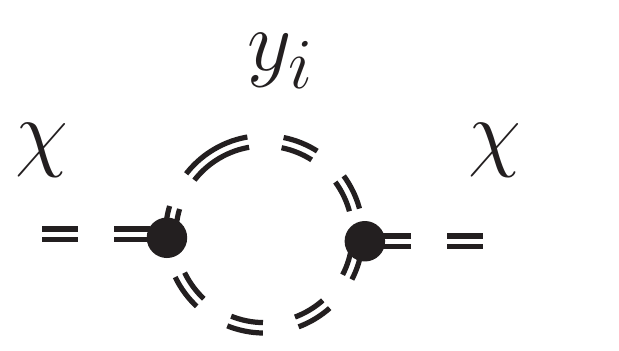} & \hspace{2em} & \includegraphics[width=4cm]{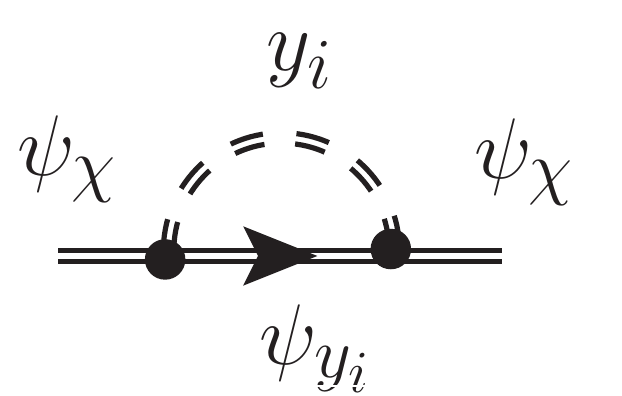} \\
     (c) & \hspace{2em} & (d)
\end{tabular}
   \caption{Feynman diagrams that account for the interaction between  heavy and light fields at one-loop,  where double lines indicate that the charged particles are dressed with the magnetic field effects. Continuous lines indicate fermionic fields, $\psi_\chi$ and $\psi_{y_i}$, and dashed lines indicate bosonic fields, $\chi$ and $y_i$. }
   \label{FDfigures}
\end{figure}

Since, we are working with a strong magnetic field, the light particles can be considered as constrained into the Lowest Landau Level (LLL), where the scalar propagator takes the form
\bea
      D^B_{y_i}(k)=\frac{2i e^{-k_\perp^2/2{q_i}B}}{k_{||}^2-{q_i}B-m_i^2+i\epsilon}.
\label{bospropLLL}
\eea
Replacing this propagator in Eq.~(\ref{LLLtad1}) and once the integration over the transverse momentum is carried out, we get
\bea
    \Sigma(p)=i\sum_{i}\frac{h^2 q_i B}{2\pi^2}\int_0^{\sqrt{q_\chi B}}    
               \frac{k_{||}dk_{||}}{k_{||}^2-q_i B-m_i^2},
\label{LLLtad2}
\eea
where the integral upper limit refers to the highest energy scale in this approach.

Performing the integration over the parallel momentum in Eq.~(\ref{LLLtad2}) by a standard procedure, we obtain
\bea
    \Sigma(p)&=&\sum_{i}\frac{h^2q_iB}{4\pi^2}\ln\left(\frac{2q_\chi B+m_i^2}{q_\chi B+m_i^2}\right)
\nonumber    \\
    &\approx&\frac{h^2(q_1+q_2)B}{4\pi^2}\ln(2).
\label{LLLtad3}
\eea

Next, the other contribution to the scalar self-energy, depicted in Fig.~\ref{FDfigures}b, comes from the interaction between one heavy scalar and two light fermion fields. In configuration space, it has the form
\bea
 -i \Sigma^B(w,v)=h^2 Tr[S_{1}^B(w,v) S_{2}^B(v,w)-\gamma^5 S_{1}^B(w,v)\gamma^5 S_{2}^B(v,w)].
\label{FDselfenergy2}
\eea
where 
\bea
S^B(x,z)=\Omega(x,z)  \tilde{S}^B(x-z)
\label{decompfer}
\eea
with $\Omega(x,z)=\exp\left[-iq\int_x^z d\xi \cdot A(\xi)\right]$, the Schwinger phase, and 
\bea
   \tilde{S}^B(x-z)=\int\frac{d^4p}{(2\pi)^4}e^{ip\cdot{(x-z)}}S^B(p)
\label{simmpartprop}
\eea
the fermion propagator symmetric part in the momentum space given in Eq.(\ref{ferpropmom}).

Note that, as the heavy field is charged, the following relation holds
\bea
    q_\chi = q_1-q_2, 
\eea
with $q_i>0$ the charge of each light particle.
 
Using the above relations in Eq.~(\ref{FDselfenergy2}), we get
\bea
     \Sigma^B(w,v)=i h^2  \Omega_{1}(w,v)\Omega_{2}(v,w) Tr[\tilde{S}_1^B(w-v) \tilde{S}_2^B(v-w)
                                         -\gamma^5 S_{1}^B(w-v)\gamma^5 S_{2}^B(v-w)],
\eea
which, in  momentum space, becomes
\bea
   \Sigma^B(k,l)&=&i h^2 \int d^4w d^4v e^{i k\cdot w}e^{-i l\cdot v}\Omega_1(w,v)\Omega_2(v,w) 
\nonumber \\
               &&\times\int \frac{d^4 p d^4 j}{(2\pi)^8} e^{-i p\cdot(w-v)}e^{-i j\cdot(v-w) }Tr[S_1^B(p)S_2^B(j)
                                     -\gamma^5S_1^B(p)\gamma^5S_2^B(j)].
\eea
Once we perform the integration over $w$ and $v$~\cite{Ryder}, we get
\bea
\Sigma^B(k,l)                
    &=&i\frac{h^24(2\pi)^6}{(q_\chi B)^2}\int \frac{d^4 p d^4 j}{(2\pi)^8}  
        \delta_{||}^{(2)}(k-p+j)\delta_{||}^{(2)}(-l+p-j) e^{-\frac{2i}{q_\chi B}\left[\epsilon_{ij}(k_i-p_i+j_i)(-l_j+p_j-j_j)\right]}      
\nonumber \\
    &&\times Tr[S_1^B(p)S^B_2(j)-\gamma^5S_1^B(p)\gamma^5S^B_2(j)],    
\eea
where $\epsilon_{ij}$ is the Levi-Civita tensor in a  2-dimensional space associated with the transverse  components. 

With the delta functions, the integral over one parallel momentum can straightforwardly be done
\bea
      \Sigma(k,l) &=&i\frac{h^24(2\pi)^4}{(q_\chi B)^2}\delta_{||}^{(2)}(k-l)
       \int \frac{d^4 p d^2 j_\perp}{(2\pi)^6} 
       e^{\frac{-2i}{[(q_\chi)B]}\left[\epsilon_{ij}(k_i-p_i+j_i)(-l_j+p_j-j_j)\right]}         
\nonumber\\
&&\times Tr[S_1^B(p)S_2^B((p-l)_{||},j_\perp)-\gamma^5S_1^B(p)\gamma^5S_2^B((p-l)_{||},j_\perp)].  
\label{selfenergyscalarfermion}
\eea
Following the same procedure as for the scalar propagator, the fermion propagator in the LLL, has the form~\cite{GusyninPLB95}
\bea
     S^{B}(p)=i \frac{e^{-\frac{p_\perp^2}{q B}}}{p_{||}^2-m^2+i\epsilon} (m+\slsh{p}_{||})(1-i\gamma_1\gamma_2).
\label{ferpropLLL}
\eea
Using this propagator in Eq.(\ref{selfenergyscalarfermion}), and once the integration over  transverse momentum and the trace over the Dirac gamma matrices are performed, we get 
\bea 
             \Sigma(k,l)&=&i16\frac{h^2q_1 q_2}{q_\chi^2} (2i)^2 (2\pi)^2\delta_{||}^{(2)}(k-l)
        e^{-\frac{(q_1+q_2) }{q_\chi^2B}(k-l)_\perp^2 }
        e^{\frac{2i}{q_\chi B} \epsilon_{ij}k_il_j} 
        \nonumber \\
      &&\times \int \frac{d^2 p_{||} }{(2\pi)^2}
       \frac{p.(p-l)_{||}}{[p_{||}^2-m_1^2+i\epsilon][(p-l)_{||}^2-m_2^2+i\epsilon]}.
\label{selfenergyscalarfermion41}
\eea
The remaining integral over the parallel momenta can be easily done resorting to the standard procedure in QFT, obtaining
\bea
     \Sigma(k,l)&&=i 16\frac{h^2q_1 q_2}{q_\chi^2} (2i)^2 (2\pi)^2\delta_{||}^{(2)}(k-l)
        e^{-\frac{(q_1+q_2) }{q_\chi^2B}(k-l)_\perp^2 }
        e^{\frac{2i}{q_\chi B} \epsilon_{ij}k_il_j} 
        \nonumber \\      
            &&\times
            \frac{i}{4\pi}
            \left\{\ln\left(\frac{m_1m_2}{\mu^2 }\right)+
             \frac{l_{||}^2-m_1^2-m_2^2}{2\sqrt{(l_{||}^2-(m_1-m_2)^2)(l_{||}^2-(m_1+m_2)^2) }}
            \right. 
      \nonumber \\
      &&\times
      \left.
        \ln\left(
          \frac{l_{||}^2-m_1^2-m_2^2+\sqrt{(l_{||}^2-(m_1-m_2)^2)(l_{||}^2-(m_1+m_2)^2)}}
                 {l_{||}^2-m_1^2-m_2^2-\sqrt{(l_{||}^2-(m_1-m_2)^2)(l_{||}^2-(m_1+m_2)^2)}}
       \right)\right\}.
\label{selfenergyfig2}
\eea
In order to match with the standard dimensions for the self-energy, which in this case are $[E^2]$, we need to invoke momentum conservation (see Appendix A) as follows
\bea
        \Sigma^B(k,l)
     &=&  (2 \pi)^4 \delta^{(2)}_{||} (k-l) \left( \frac{1}{q_\chi B}\right)
       \widetilde\Sigma\left(l,k_\perp\right),    
\label{genresself}
\eea
where the factor that involves the magnetic field emphasises the momentum conservation in the transverse direction. In such a way, from Eq.(\ref{selfenergyfig2}), we get 
\bea
     \widetilde{\Sigma}(k,l_\perp)&&=4\frac{h^2q_1 q_2 q_\chi B}{q_\chi^2 \pi^3}  
        e^{-\frac{(q_1+q_2) }{q_\chi^2B}(k-l)_\perp^2 }
        e^{\frac{2i}{q_\chi B} \epsilon_{ij}k_il_j} 
        \nonumber \\      
            &&\times
\left\{\ln\left(\frac{m_1m_2}{\mu^2 }\right)+
             \frac{l_{||}^2-m_1^2-m_2^2}{2\sqrt{(l_{||}^2-(m_1-m_2)^2)(l_{||}^2-(m_1+m_2)^2) }}
            \right. 
      \nonumber \\
      &&\times
      \left.
        \ln\left(
          \frac{l_{||}^2-m_1^2-m_2^2+\sqrt{(l_{||}^2-(m_1-m_2)^2)(l_{||}^2-(m_1+m_2)^2)}}
                 {l_{||}^2-m_1^2-m_2^2-\sqrt{(l_{||}^2-(m_1-m_2)^2)(l_{||}^2-(m_1+m_2)^2)}}
       \right)\right\}.
\label{resultselffromfermions}       
\eea
In the above equation, strictly speaking, we do not have a Dirac delta function that accounts for the transverse momentum conservation of the heavy field, nevertheless, as both $l_\perp$ and $k_\perp$ are weighted by the magnetic field, they are of the same order of magnitude, which is negligible within this approximation. With this in mind, we hereafter take equal $l$ and $k$, and Eq.~(\ref{resultselffromfermions}) reduces to
\bea
     \widetilde{\Sigma}(l)&&=\frac{4}{\pi^3}\frac{h^2q_1 q_2  B}{q_\chi }  
        \nonumber \\      
            &&\times
\left\{\ln\left(\frac{m_1m_2}{\mu^2 }\right)+
             \frac{l_{||}^2-m_1^2-m_2^2}{2\sqrt{(l_{||}^2-(m_1-m_2)^2)(l_{||}^2-(m_1+m_2)^2) }}
            \right. 
      \nonumber \\
      &&\times
      \left.
        \ln\left(
          \frac{l_{||}^2-m_1^2-m_2^2+\sqrt{(l_{||}^2-(m_1-m_2)^2)(l_{||}^2-(m_1+m_2)^2)}}
                 {l_{||}^2-m_1^2-m_2^2-\sqrt{(l_{||}^2-(m_1-m_2)^2)(l_{||}^2-(m_1+m_2)^2)}}
       \right)\right\},
\label{resultselffromfermionstaylorx1}       
\eea
where $\mu$ is an ultraviolet cuttoff, that can be related with the magnetic field $(\mu^2\propto q_1B)$. As the main contribution to the mass comes from the region $l^2_{||}\lesssim q_1 B$, Eq.~(\ref{resultselffromfermionstaylorx1}) becomes
\bea
     \widetilde{\Sigma}(l)&&=-\frac{4}{\pi^3}\frac{h^2q_1 q_2  B}{q_\chi }  
       \ln\left(\frac{q_1B }{m_1m_2}\right).
\label{resultselffromfermionstaylorx2}       
\eea
The interaction between one heavy and two light scalar fields, Feynman diagram in Fig.~\ref{FDfigures}c, does not contribute to the scalar self-energy as shown in \cite{HM},  in this way we ignore it.

Taking into account both contributions, Eq.~(\ref{resultselffromfermionstaylorx2}) and  Eq.~(\ref{LLLtad3}), the total magnetic effect to the heavy boson mass reduces to 
\bea
     m_\chi^2(B)\approx M^2-\frac{4}{\pi^3}\frac{h^2q_1 q_2  B}{q_\chi }  
       \ln\left(\frac{q_1B }{m_1m_2}\right),
\label{eq:bosonselfenergy}
\eea
with $M^2=2g^2\phi^2$.

\subsubsection{Fermion self-energy dressed with magnetic field effects}

In configuration space the self-energy of the fermionic heavy field, interacting with one fermion and one scalar light fields,  shown in Fig.~\ref{FDfigures}d, has the form
\bea
  -i\Sigma^B(w,v)&=&-2h^2  D_{y}^B(w,v) S_{\psi_y}^B(v,w)
  \nonumber \\
                 &=&-2h^2  \Omega_{y}(w,v)\Omega_{\psi_y}(v,w) \tilde{D}_y^B(w-v) \tilde{S}_{\psi_y}^B(v-w),  
\eea
where we have decomposed the boson propagator in the same way as the fermion propagator in Eq.(\ref{decompfer}).  

In  momentum space, the above equation becomes
\bea
       \Sigma^B(k,l)=-2ih^2 \int d^4w d^4v e^{i k\cdot w}e^{-i l\cdot v}\Omega_1(w,v)\Omega_2(v,w) 
                         \int \frac{d^4 p d^4 j}{(2\pi)^8} e^{-i p\cdot(w-v)}e^{-i j\cdot(v-w) }D_y^B(p)S_{\psi_y}^B(j).
\eea
Once we perform the integration over $w$ and $v$~\cite{Ryder}, we get
\bea
       \Sigma^B(k,l)
    =-i\frac{2h^24(2\pi)^6}{(q_\chi B)^2}\int \frac{d^4 p d^4 j}{(2\pi)^8}D_y^B(p)S^B_{\psi_y}(j)   
        \delta_{||}^{(2)}(k-p+j)\delta_{||}^{(2)}(-l+p-j) e^{-\frac{2i}{q_\chi B}\left[\epsilon_{ij}(k_i-p_i+j_i)(-l_j+p_j-j_j)\right]}.     
\eea

With the delta functions, the integral over one parallel momentum can straightforwardly be done
\bea
       \Sigma(k,l) =-i\frac{128\pi^4h^2}{(q_\chi B)^2}\delta_{||}^{(2)}(k-l)
       \int \frac{d^4 p d^2 j_\perp}{(2\pi)^6}D_y^B(p)S_{\psi_y}^B((p-l)_{||},j_\perp)  
       e^{\frac{-2i}{[(q_\chi)B]}\left[\epsilon_{ij}(k_i-p_i+j_i)(-l_j+p_j-j_j)\right]}. 
       \label{eq51}
\eea
Using the LLL boson and fermion propagators, Eqs.~(\ref{bospropLLL}) and (\ref{ferpropLLL}), in Eq.~(\ref{eq51}), and once the integration over transverse momentum is carried out, we arrive at
\bea
       \Sigma(k,l)&=&i\frac{64\pi^2 h^2q_1 q_2}{q_\chi^2} \delta_{||}^{(2)}(k-l)
        e^{-\frac{(q_1+q_2) }{q_\chi^2B}(k-l)_\perp^2 }
        e^{\frac{2i}{q_\chi B} \epsilon_{ij}k_il_j} 
        \nonumber \\
      &&\times \int \frac{d^2 p_{||} }{(2\pi)^2}
       \frac{(\slsh{p}_{||}-\slsh{l}_{||}+m_2)(1-i\gamma_1\gamma_2)}{[p_{||}^2-q_yB-m_1^2+i\epsilon][(p-l)_{||}^2-m_2^2+i\epsilon]}.
\eea
The remaining integral over the parallel momenta can be easily done resorting to the standard procedure on QFT, obtaining
\bea
      \Sigma(k,l)&=&-\frac{16\pi h^2q_1 q_2}{q_\chi^2}  \delta_{||}^{(2)}(k-l)
        e^{-\frac{(q_1+q_2) }{q_\chi^2B}(k-l)_\perp^2 }
        e^{\frac{2i}{q_\chi B} \epsilon_{ij}k_il_j} 
        \nonumber \\
      && \left\{
               \frac{\slsh{l_{||}}}{2l_{||}^2}\ln\left(\frac{m_2^2}{m_1^2+q_1B}\right)
                +\left(m_2+\frac{\slsh{l}_{||}(l_{||}^2+m_2^2-m_1^2)}{2l_{||}^2}\right)
             \right.
      \frac{i}{\sqrt{4l_{||}^2 m_2^2-(-l_{||}^2+m_1^2-m_2^2+q_1B)^2 }}
      \nonumber \\
      &&\times\left.
        \ln\left(
          \frac{l_{||}^2-m_1^2-m_2^2-q_1B+i\sqrt{4l_{||}^2 m_2^2-(-l_{||}^2+m_1^2-m_2^2+q_1B)^2 }}
                 {l_{||}^2-m_1^2-m_2^2-q_1B-i\sqrt{4l_{||}^2 m_2^2-(-l_{||}^2+m_1^2-m_2^2+q_1B)^2 }}
       \right)\right\}(1-i\gamma_1\gamma_2).
\eea
Once we employ Eq.~(\ref{genresself}), and bearing in mind the argument above Eq.(\ref{resultselffromfermionstaylorx1}),  we get 
\bea
       \tilde{\Sigma}(l)&=&-\frac{1}{\pi^3}\frac{h^2q_1 q_2 B}{2 q_\chi l_{||}^2}
            \left\{
               \slsh{l_{||}}\ln\left(\frac{m_2^2}{m_1^2+q_1B}\right)
                +
             \right.
      \frac{2l_{||}^2m_2+\slsh{l}_{||}(l_{||}^2+m_2^2-m_1^2)}{\sqrt{\left(l_{||}^2-(m_2+\sqrt{m_1^2+q_1B})^2\right)\left(l_{||}^2-(m_2-\sqrt{m_1^2+q_1B})^2\right)}}
      \nonumber \\
      &&\hspace{-5em}\times\left.
        \ln\left(
          \frac{l_{||}^2-m_1^2-m_2^2-q_1B-\sqrt{\left(l_{||}^2-(m_2+\sqrt{m_1^2+q_1B})^2\right)\left(l_{||}^2-(m_2-\sqrt{m_1^2+q_1B})^2\right)}}
                 {l_{||}^2-m_1^2-m_2^2-q_1B+\sqrt{\left(l_{||}^2-(m_2+\sqrt{m_1^2+q_1B})^2\right)\left(l_{||}^2-(m_2-\sqrt{m_1^2+q_1B})^2\right)}}
       \right)\right\}(1-i\gamma_1\gamma_2).
\label{resselfferm}       
\eea
In order to calculate the magnetic mass contribution, following the procedure sketched in the Appendix B in Ref.\cite{Warmus}, let us write the propagator dressed with the magnetic field through  the quantum correction as follows
\bea
    S^{-1}_{B}&=&\slsh{l}-M -\tilde{\Sigma}^B 
\nonumber \\
     &=&\left(\slsh{l} - M -\tilde{\Sigma}^B \right) \Delta(+)  +\left(\slsh{l} - M -\tilde{\Sigma}^B \right) \Delta(-), 
\label{propsigma}               
\eea
with  $\Delta(\pm) \equiv \frac{1}{2}\left( 1\pm \Sigma_3\right)$, the spin projectors along the magnetic field direction. Once we replace Eq.~(\ref{resselfferm}) in Eq.(\ref{propsigma}), the above equation can be rewritten as
\bea
    S^{-1}_{B}&=&\left(\slsh{l} - M^+  \right) \Delta(+)+\left(\slsh{A} - M^-\right) \Delta(-)  
\label{propawithquantumcorr}
\eea
where the asymmetry in the two terms rises as consequence of the interaction between the strong magnetic field and the spin particle.  The $A^\mu$ components and the $M^\pm$ are defined as
\bea
A^0&=&l^0 +l^0\frac{h^2q_1 q_2 B}{\pi^3 q_\chi l_{||}^2}
            \big\{
               \ln\left(\frac{m_2^2}{m_1^2+q_1B}\right)
                +
      \frac{(l_{||}^2+m_2^2-m_1^2)}{\sqrt{\left(l_{||}^2-(m_2+\sqrt{m_1^2+q_1B})^2\right)\left(l_{||}^2-(m_2-\sqrt{m_1^2+q_1B})^2\right)}}
      \nonumber \\
      &&\times
        \ln\left(
          \frac{l_{||}^2-m_1^2-m_2^2-q_1B-\sqrt{\left(l_{||}^2-(m_2+\sqrt{m_1^2+q_1B})^2\right)\left(l_{||}^2-(m_2-\sqrt{m_1^2+q_1B})^2\right)}}
                 {l_{||}^2-m_1^2-m_2^2-q_1B+\sqrt{\left(l_{||}^2-(m_2+\sqrt{m_1^2+q_1B})^2\right)\left(l_{||}^2-(m_2-\sqrt{m_1^2+q_1B})^2\right)}}
       \right)\big\},
\nonumber \\
A^1&=& l^1,\nonumber \\
A^2&=& l^2, \nonumber \\
A^3&=&l^3 +l^3\frac{h^2q_1 q_2 B}{\pi^3 q_\chi l_{||}^2}
            \big\{
               \ln\left(\frac{m_2^2}{m_1^2+q_1B}\right)
                +
      \frac{(l_{||}^2+m_2^2-m_1^2)}{\sqrt{\left(l_{||}^2-(m_2+\sqrt{m_1^2+q_1B})^2\right)\left(l_{||}^2-(m_2-\sqrt{m_1^2+q_1B})^2\right)}}
      \nonumber \\
      &&\times
        \ln\left(
          \frac{l_{||}^2-m_1^2-m_2^2-q_1B-\sqrt{\left(l_{||}^2-(m_2+\sqrt{m_1^2+q_1B})^2\right)\left(l_{||}^2-(m_2-\sqrt{m_1^2+q_1B})^2\right)}}
                 {l_{||}^2-m_1^2-m_2^2-q_1B+\sqrt{\left(l_{||}^2-(m_2+\sqrt{m_1^2+q_1B})^2\right)\left(l_{||}^2-(m_2-\sqrt{m_1^2+q_1B})^2\right)}}
       \right)\big\},
\nonumber \\
M^+&=& M,       
\nonumber \\                      
M^- &=& M - \frac{2m_2h^2q_1 q_2 B}{\pi^3 q_\chi}
      \frac{
                    \ln\left(
          \frac{l_{||}^2-m_1^2-m_2^2-q_1B-\sqrt{\left(l_{||}^2-(m_2+\sqrt{m_1^2+q_1B})^2\right)\left(l_{||}^2-(m_2-\sqrt{m_1^2+q_1B})^2\right)}}
                 {l_{||}^2-m_1^2-m_2^2-q_1B+\sqrt{\left(l_{||}^2-(m_2+\sqrt{m_1^2+q_1B})^2\right)\left(l_{||}^2-(m_2-\sqrt{m_1^2+q_1B})^2\right)}}
       \right)
      }{\sqrt{\left(l_{||}^2-(m_2+\sqrt{m_1^2+q_1B})^2\right)\left(l_{||}^2-(m_2-\sqrt{m_1^2+q_1B})^2\right)}}.
      \nonumber \\
\eea
Now, replacing Eq.(\ref{propawithquantumcorr}) in the effective potential given in Eq.(\ref{potfer1loop}), the magnetic correction to the heavy fermion mass can be identified, in the strong field limit, as
\bea
     m_{\psi_\chi}^2(B)=M^2+\frac{h^2 q_2 }{\pi^3 q_\chi }\left\{q_1 B \ln\left(\frac{q_1 B}{m_2^2}\right)+m_1^2-(l_{||}^2-m_1^2+2m_{\psi_\chi}m_2+m_2^2)\ln\left(\frac{q_1 B}{m_2^2}\right)\right\},
\eea
where we have kept terms up to $\mathcal{O}(h^2)$, consistent with the perturbative expansion  done in this work. Since the main contribution to the mass comes from the $l^2_{||}\lesssim q_1 B$ region, then the above equation reduces to
\bea
    m_{\psi_\chi}^2(B)=M^2+\frac{h^2q_1 q_2 B}{\pi^3 q_\chi }\ln\left(\frac{q_1 B}{m_2^2}\right).
\label{eq:fermionselfenergy}   
\eea
With the above result we conclude the magnetic contribution to the heavy particle  masses. By comparing Eqs.~(\ref{eq:bosonselfenergy}) and (\ref{eq:fermionselfenergy}), we note that the soft-SUSY breaking term, introduced by $M_s^2$ in the boson mass, has an additional source for symmetry breaking coming from the magnetic field. 

\subsection{ Heavy charged scalar decay width into two light charged fermions}\label{subsecdecayB}

Since the main ingredient in the warm inflation scenario is a dissipation term coming from the inflaton decay that dominates over the one generated by the universe expansion, in this subsection we compute the effect of the magnetic field on the decay process of a scalar boson to charged fermions, in order to show the trend of the magnetic contribution. The decay width of a scalar particle to two bosons (Fig.~\ref{FDfigures}(c)) can easily be estimated following the same calculation scheme, although this diagram is subdominant with respect to the one in Fig.~\ref{FDfigures}(b). On another hand, the study of the decay, in the presence of a magnetic field, of a fermion into a fermion and a scalar (Fig.~\ref{FDfigures}(d)), where all of them are charged, is more involved and will be worked out elsewhere.

In order to calculate the charged scalar particle decay width we shall make use of the optical theorem, which relates the self-energy imaginary part with the decay width, as follows 
\bea
     \Gamma(l)=-\frac{\text{Im}(\Sigma(l))}{l_0}.
\label{decay1}
\eea
Since the main contribution to scalar self-energy comes from Fig.\ref{FDfigures}(b), let us start by rewritting Eq.~(\ref{selfenergyscalarfermion41}), as 
\bea
   \tilde{\Sigma}(l)&=&-i\frac{16}{\pi^2}h^2\frac{q_1 q_2}{q_\chi}  B\
      \int \frac{d^2 p_{||} }{(2\pi)^2}
       \frac{p.(p-l)_{||}}{[p_{||}^2-m_1^2+i\epsilon][(p-l)_{||}^2-m_2^2+i\epsilon]},
\label{decay2}
\eea
where we have used the ansatz in Eq.~(\ref{genresself}) and kept the integration over the internal momentum. Note that, in the low transverse momentum approximation, we can safely  ignore the exponential factors due to they are suppressed by the magnetic field.

Next, with the help of the identity
\bea
    \frac{1}{A}=-i \int_0^\infty ds \ e^{i s A},
\eea
where  $\text{Im}(A)>0$ is assumed, we rewrite the denominators in Eq.(\ref{decay2}), getting 
\bea
   \tilde{\Sigma}(l)&=&i\frac{16}{\pi^2}h^2\frac{q_1 q_2}{q_\chi}  B\
      \int \frac{d^2 p_{||} }{(2\pi)^2}\ ds_1ds_2 \
       p.(p-l)_{||}\ e^{is_1[p_{||}^2-m_1^2+i\epsilon]}e^{is_2[(p-l)_{||}^2-m_2^2+i\epsilon]}.
\label{decay3}
\eea
Now, by using the change of variables
\bea
   s_1=\frac{s}{2}(1+v) \mbox{\ \ and \ \ } s_2=\frac{s}{2}(1-v),
\eea
and performing the Gaussian integrals over the $p$ momentum, we get 
\bea
   \tilde{\Sigma}(l)&=&i\frac{16}{\pi^2}h^2\frac{q_1 q_2}{q_\chi}  B\
      \int_{-1}^1  dv\int_0^\infty ds \ \frac{1}{8\pi}\left(\frac{i}{s}-\frac{1-v^2}{4}l^2\right)
        e^{is[\frac{1-v^2}{4}l_{||}^2-m_1^2\frac{(1+v)}{2}-m_2^2\frac{(1-v)}{2}+i\epsilon]},
\label{decay4}
\eea
where the $i\epsilon$ was omitted for the sake of simplicity.
Note that the first term, in Eq.(\ref{decay4}), becomes divergent in the region $s\rightarrow 0$.  In order to isolate the main divergence in the above equation, we integrate by parts over $v$, and we obtain 
\bea
   \tilde{\Sigma}(l)&=&i\frac{16}{\pi^2}h^2\frac{q_1 q_2}{q_\chi}  B\ \left\{
      \int_0^\infty \frac{i ds}{8\pi s}\left(e^{-is(m_1^2-i\epsilon)}+e^{-is(m_2^2-i\epsilon)}\right)\right.
\nonumber \\
    &&\hspace{1cm}-\left. \int_{-1}^1  dv\int_0^\infty ds \ \frac{1}{32\pi}\left((1+v^2)l_{||}^2+2v(m_1^2-m_2^2)\right)
        e^{is[\frac{1-v^2}{4}l_{||}^2-m_1^2\frac{(1+v)}{2}-m_2^2\frac{(1-v)}{2}+i\epsilon]}\right\}.     
\label{decay5}
\eea
It is not difficult to see that the first term in the above equation is a real quantity that does not contribute to the self-energy imaginary part and thus we shall ignore it. For the second term, once the integration over $s$ is carried out, we arrive at
\bea
    \Sigma(l)&=&\frac{16}{\pi^2}h^2\frac{q_1 q_2}{q_\chi}  B 
        \int_{-1}^1  dv \ \frac{1}{32\pi}\frac{(1+v^2)l_{||}^2+2v(m_1^2-m_2^2)}
        {\frac{1-v^2}{4}l_{||}^2-m_1^2\frac{(1+v)}{2}-m_2^2\frac{(1-v)}{2}+i\epsilon}. 
\label{decay6} 
\eea
Now, with the help of the identity
\bea
     \lim_{\epsilon\rightarrow0}\left(\mbox{Im}\left(\frac{1}{x+i\epsilon}\right)\right)=-\pi \delta(x),
\eea
the self-energy imaginary part of Eq.(\ref{decay6}), reads
\bea
   \mbox{Im}(\Sigma^B(l))&=&-\frac{h^2}{2\pi^2}\frac{q_1 q_2}{q_\chi}  B 
        \int_{-1}^1  dv \ 4\left((1+v^2)l_{||}^2+2v(m_1^2-m_2^2)\right)\delta\left(
        {(1-v^2)l_{||}^2-2m_1^2(1+v)-2m_2^2(1-v)}\right). 
\nonumber \\
    &=&-\frac{h^2}{\pi^2}\frac{q_1 q_2}{q_\chi}  B \frac{2(l_{||}^2-m_1^2-m_2^2)}{\sqrt{(l_{||}^2-(m_1+m_2)^2)(l_{||}^2-(m_1-m_2)^2)}}\theta(l_{||}^2-(m_1+m_2)^2),
\eea
where the $\theta$ function emerges as a consequence that $v\in [-1,1]$.

With the above at hand, then the decay width for this charged particle reads
\bea
      \Gamma^B(l)=\frac{2h^2}{\pi^2}\frac{q_1 q_2}{q_\chi}  B \frac{l_{||}^2-m_1^2-m_2^2}{l_0\sqrt{(l_{||}^2-(m_1+m_2)^2)(l_{||}^2-(m_1-m_2)^2)}}\theta(l_{||}^2-(m_1+m_2)^2),
\label{decay7}      
\eea
showing that the decay process grows linearly with the magnetic field.

In order to extract the magnetic field effect on the decay width, we repeat the above procedure in the case without any external magnetic field, getting
\bea
      \Gamma(l)=\frac{h^2}{4\pi}\frac{(l^2-m_1^2-m_2^2)}{l_0 l^2 }\sqrt{(l^2-(m_1+m_2)^2)(l^2-(m_1-m_2)^2)}\theta(l^2-(m_1+m_2)^2).
\label{decay8}
\eea
In Fig.~(\ref{result2}), we plot the ratio $ \Gamma^B(l)/ \Gamma(l)$, as a function of the magnetic field strength, with $M_s/\phi_0=0.05$, $m_1/\phi_0=10^{-3}$, $m_2/\phi_0=5\times10^{-3}$,  $g=0.1$, for a range allowed by the energy hierarchy scale.  
\begin{figure}[h!]
	\includegraphics[width=10cm]{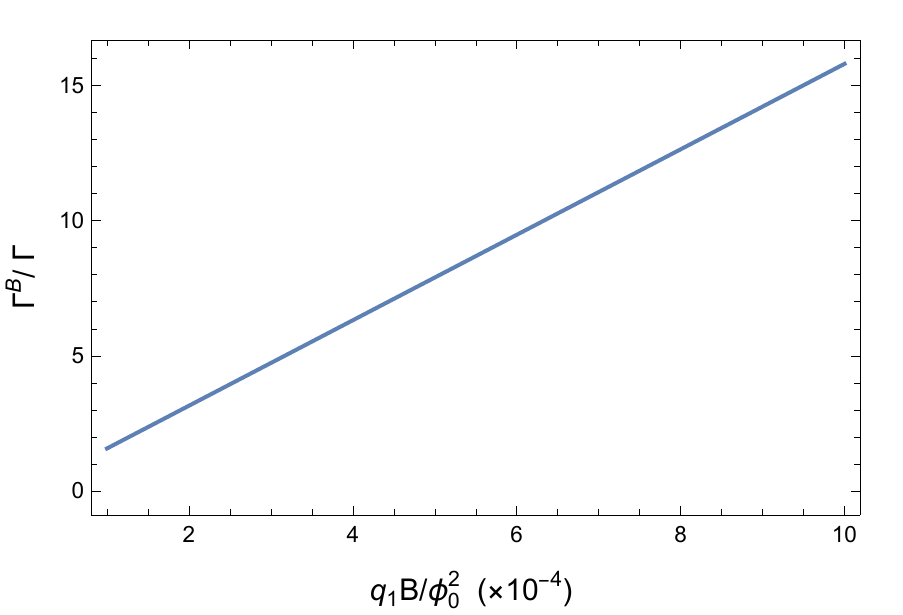}
	\caption{Decay width ratio $\Gamma^B/\Gamma$, Eqs.~(\ref{decay7}-\ref{decay8}), of a heavy charged scalar into two  light charged fermions, for different magnetic field strengths, for $M_s/\phi_0=0.05$, $m_1/\phi_0=10^{-3}$, $m_2/\phi_0=5\times10^{-3}$,  $g=0.1$. The magnetic field enhances this process.}
	\label{result2}
\end{figure}

In this figure, it is clear that the presence of a strong magnetic field enhances the heavy  scalar decay width into a pair of charged fermions. This result seems to reinforce one required condition within a warm inflationary scenario, which is that the inflaton interaction with other fields becomes important during the inflationary stage.  This statement is done by assuming that the relation between the heavy particle decay $\chi$ and the inflaton field $\phi$ in the absence of an external magnetic field, given by  \cite{HM}
\bea
     \Gamma_\phi=\frac{g^4\phi^4\Gamma_\chi}{2\pi(m_\chi^2+\Gamma_\chi^2)[2m_\chi(m_\chi^2+\Gamma_\chi^2)^{1/2}+2m_\chi^2]^{1/2}},
\label{infldecay}
\eea
 holds in our case. Of course, Eq.~(\ref{infldecay}) must be modified taking into account that the external heavy particles are affected by the external magnetic field. This is an involved calculation, that needs a careful analysis. In the literature different results have been reported for the decay process in the presence of an external magnetic field: some authors have obtained that the magnetic field enhances the decay process, however other authors get an opposite effect. All these works seem to indicate that the behavior of decay process with the magnetic field strongly depends on the kinematics and particles spin involved in the initial and final states, as well as the external magnetic field strength (see {\it e.g.}~\cite{JaberPiccinelliSanchez} and references therein).   The specific magnetic field effect on the inflaton dissipation coefficient will be reported elsewhere.

\section{Results}\label{sec5}

\subsection{Effective potential} 

In this section, we put together all the magnetic contributions to the inflaton one-loop effective potential, including thermal and magnetic effects, getting
\bea
    V(\phi,T,B)=-\frac{\pi}{90}g^*T^4+\mathcal{V}^1(\phi,B),
\eea
where $\mathcal{V}^1(\phi,B)$ is the potential in Eq.(\ref{potentialfull}) including the magnetic masses.

In order to quantify the magnetic field effect on the effective potential, let us define 
\bea
     \Delta V(\phi,T,B) &\equiv& \frac{V^{(1)}(\phi,T,B)-V^{(1)}(0,T,B)}{V^{(1)}(0,0,0)}+1\nonumber \\ &=&\Delta \mathcal{V}(\phi,B)
     ,
\label{forplots}
\eea
where we compare the effective potential depth in the presence of the magnetic field with the case without magnetic field. Note that, in the physical scenario considered in this work, Eq.~(\ref{forplots}) does not depend on temperature.

In Fig.~(\ref{result1}), we plot Eq.(\ref{forplots}) as a function of $\phi/\phi_0$ for different magnetic field strengths. Taking into account the limits imposed on the coupling constants by the slow-roll conditions and the constraints from density perturbations \cite{HM}, the values we have chosen in the effective potential are in concordance with the hierarchy scale  used in the calculations: $M_s/\phi_0=0.05$, $m_1/\phi_0=10^{-3}$, $m_2/\phi_0=5\times10^{-3}$,  $g=0.1$ and $h=0.1$, resulting in $M_{\chi}/\phi_0=0.15$, $M_{\psi_\chi}/\phi_0=0.14$ and $\Lambda/\phi_0=0.3$.

As can be seen, the effect of the magnetic field on the effective potential is to make it less steep, preserving the slow-roll conditions. \begin{figure}[h!]
	\includegraphics[width=10cm]{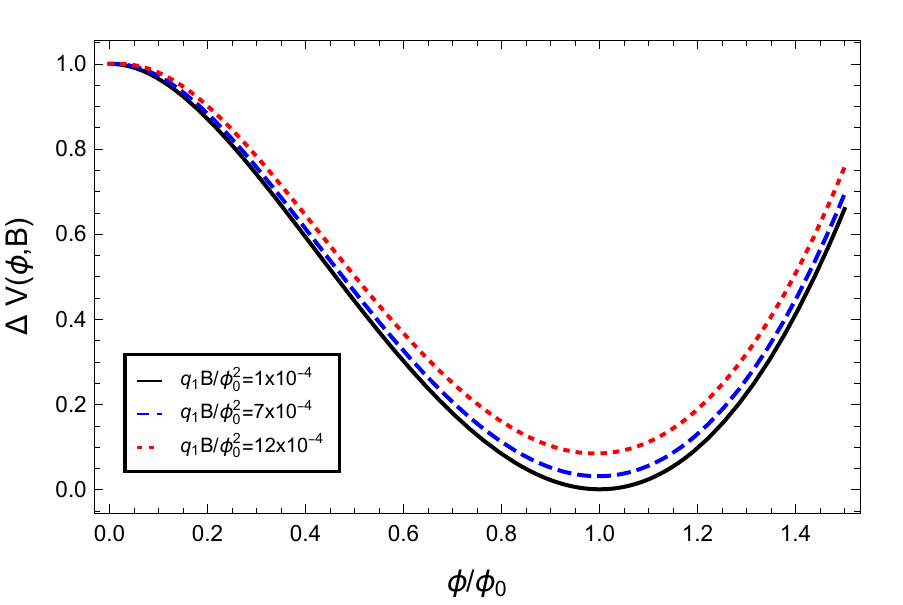}
	\caption{Effective potential normalized by $\mathcal{V}(0,0)$, Eq.~(\ref{forplots}),  for different magnetic field strengths for $M_s/\phi_0=0.05$, $m_1/\phi_0=10^{-3}$, $m_2/\phi_0=5\times10^{-3}$,  $g=0.1$ and $h=0.1$.}
	\label{result1}
\end{figure}

\subsection{Slow-roll parameters}
The slow-roll conditions can be verified through the parameters
\bea
    \epsilon=\frac{m_P^2}{16\pi}\left(\frac{V'}{V}\right)^2   \mbox{   and  }  \eta=\frac{m_P^2}{8\pi}\left(\frac{V''}{V}\right),
\label{}
\eea
which, in the case of warm inflation, are bounded by~\cite{microphysical, Bereraeta}
\bea
\epsilon <  1+ \frac{\Gamma_\phi}{3H}  \mbox{\hspace{1cm} and \hspace{1cm}} 1 < \eta < 1+ \frac{\Gamma_\phi}{3H}.
\label{boundsslowroll}
\eea

For our model, in Fig.~\ref{result4}~(a), we plot $\epsilon$ as a function of $\phi/\phi_0$, for different magnetic field strengths, showing that the magnetic field effect keeps this slow-roll parameter below the unit for a wider range of $\phi/\phi_0$. As shown in Fig.~\ref{result4}~(b), the $\phi/\phi_0$-values where $\epsilon = 1$ depend on the magnetic field strength, supporting the idea that the magnetic field effect preserves the potential flatness. Taking as a reference point the value  $\epsilon =1$,  which sets the end of inflation in supercooled inflationary models, the standard relation between the e-folds number and $\epsilon$-parameter, suggests that within this model 60 e-folds inflationary expansion can be achieved. An accurate estimation for the e-fold number must be done with the help of Eq.(2.35) in Ref.~\cite{microphysical}, in which the $\Gamma_\phi$ plays an important role. 
\begin{figure}[h!]
\begin{tabular}{cc}
	\includegraphics[width=8cm]{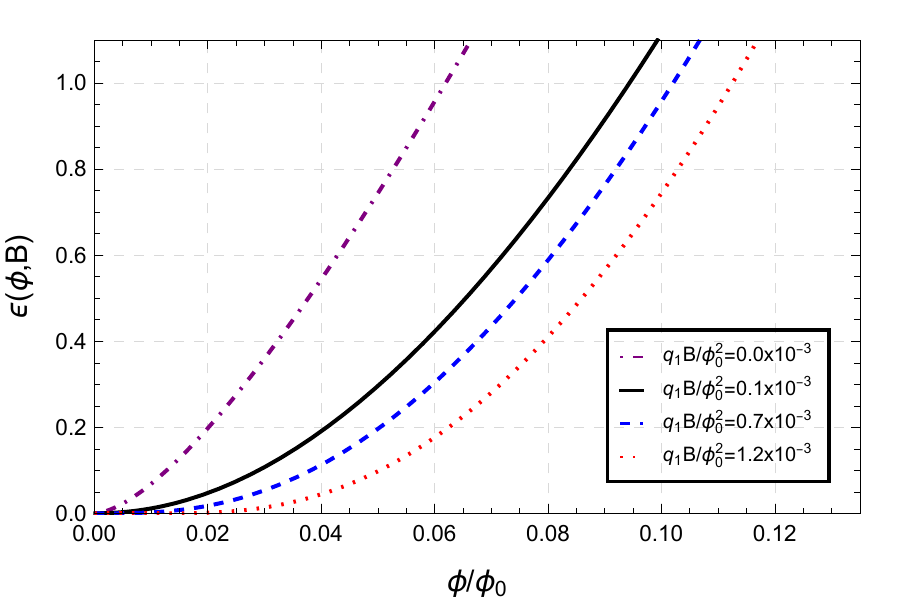}
&
	\includegraphics[width=8cm]{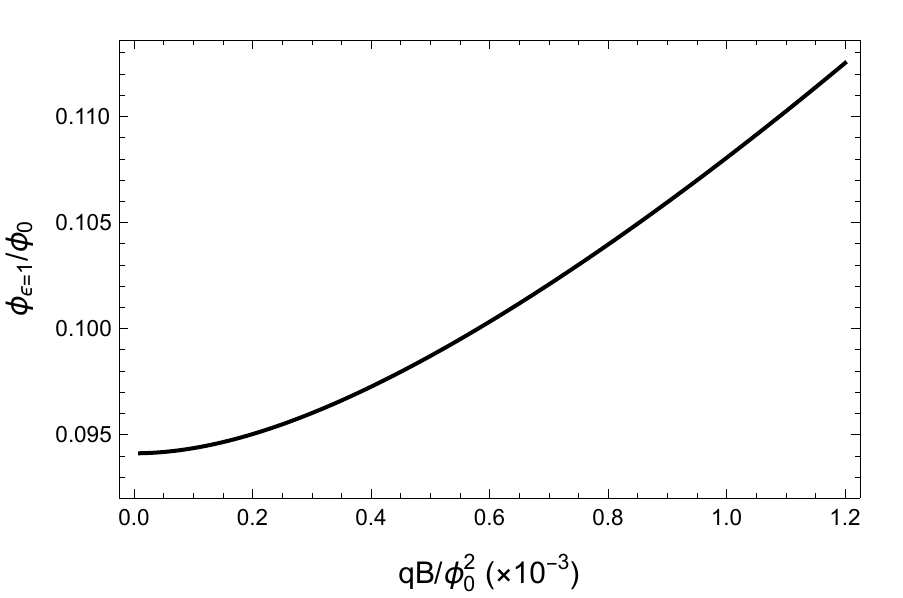}\\
		(a) & (b) \\
\end{tabular}
	\caption{(a) Slow-roll parameter $\epsilon$ for different magnetic field strengths and (b) $\phi_{\epsilon=1}/\phi_0$ as a function of the magnetic field, both with $M_s/\phi_0=0.05$, $m_1/\phi_0=10^{-3}$, $m_2/\phi_0=5\times10^{-3}$,  $g=0.1$ and $h=0.1$.}
	\label{result4}
\end{figure}

In the  $\eta$-case, its value strongly depends on the stage in the inflationary process. In order to get an idea about the effect of the magnetic field on this parameter, let us consider two different stages (a) at $\epsilon=1$, keeping in mind that for warm inflation this value does not represent the end of inflationary expansion~\cite{BereraMossHall}, and (b) an earlier inflationary stage where $\epsilon<1$. In these neighborhoods, the $\eta$-parameter is plotted as a function of $\phi/\phi_0$ in Fig. 5, for different values of the magnetic field strength, showing that its effect is to enhance the $\eta$-value. This seems to favor the warm inflation scenario $(\eta >1)$, nonetheless, the upper bound in Eq.~(\ref{boundsslowroll}) needs to be checked. As we have mentioned by the end of Sec.~\ref{subsecdecayB}, the calculation for $\Gamma_\phi$ is a work in progress and is needed to verify that the upper limit in the $\eta$-parameter be fulfilled. 

\begin{figure}
\begin{tabular}{cc}
	\includegraphics[width=8cm]{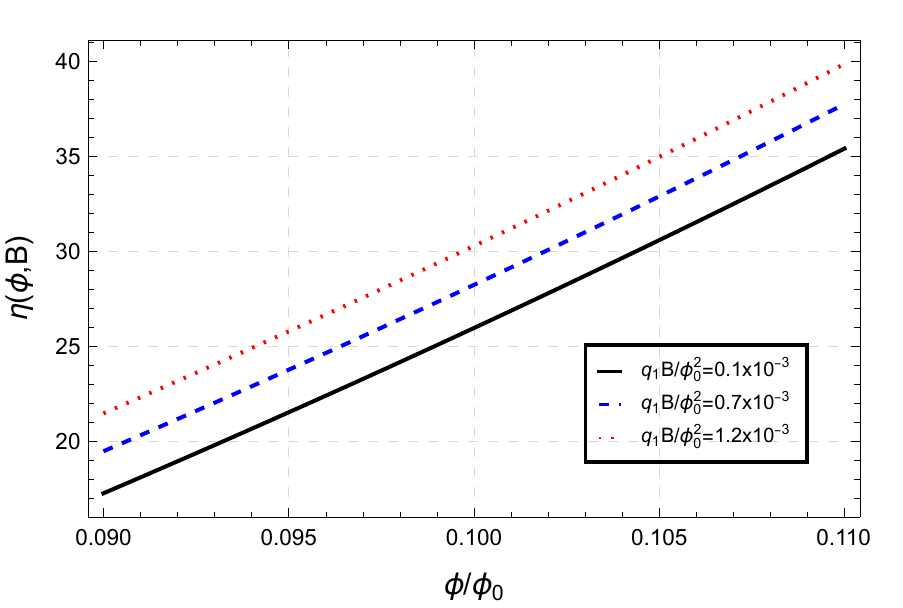}
&
	\includegraphics[width=8cm]{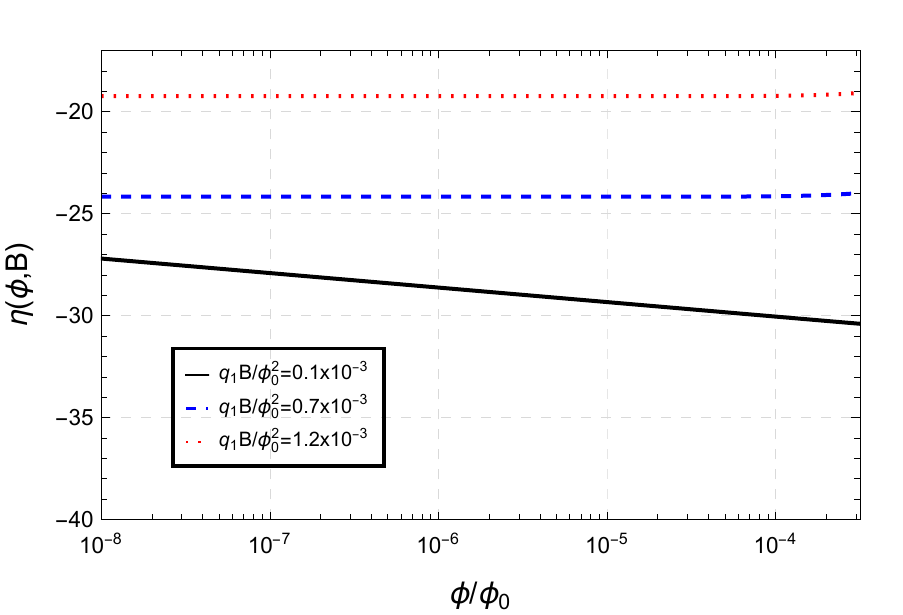} \\
	(a) & (b) \\
\end{tabular}
\caption{$\eta$ as a function of $\phi/\phi_0$ for three values of the magnetic field strength, (a) in the neighborhood of $\phi/\phi_0$ for which $\epsilon=1$ and (b) for an earlier inflationary stage. In both cases, $M_s/\phi_0=0.05$, $m_1/\phi_0=10^{-3}$, $m_2/\phi_0=5\times10^{-3}$,  $g=0.1$ and $h=0.1$.}
\end{figure}

\section{Conclusions}\label{sec6}

In this paper we have studied the effects that a possible primordial magnetic field can have on the inflation’s potential, taking as the underlying model a warm inflation scenario and considering that all fields interacting with the inflaton field are charged. The model is based on global supersymmetry and a coupling between the inflaton and heavy intermediate superfields which are in turn coupled to light particles. In this context, since we have considered a possible effect on the heavy field sector, then we worked in the strong magnetic field approximation for the light fields. This limit was taken when computing the light particles contribution to the heavy sector self-energy, but the expression for the effective potential, up to one-loop, is exact. We studied its behavior, choosing coupling constants and masses in concordance with the hierarchy scale used in the calculations. 

This work is an extension of a previous one in which the weak magnetic field limit was considered and only the light sector was charged. As in our previous study, here we found that the magnetic field effect on the effective potential is to make it less steep as compared with the vacuum case, showing that magnetic fields do not spoil the inflationary process. This statement is supported by the behavior shown by the slow-roll $\epsilon$-parameter as a function of the magnetic field. In the case of the $\eta$-parameter, we have incorporated the magnetic field effect on it and we have checked that the lower limit for this parameter $(1<\eta)$ is fulfilled, however, we could not reach a firm conclusion about the upper limit $(\eta<1+\Gamma_\phi/3H)$, since the latter depends directly on the inflaton decay rate that needs to be analyzed carefully.  In this work, we further explore the viability of this scenario by estimating the magnetic field effect on the heavy charged scalar decay width, finding an enhancement on this process. A complete study of the decay process of the particles involved in this model under the influence of the external magnetic field is in progress and will be presented elsewhere.

Our findings could be relevant in the scrutiny of the role played by magnetic fields on cosmological events, since there are good chances that they were present during the early stages of the universe, where phase transitions provided suitable conditions for their generation.

\appendix

\section{On the self-energy physical dimensions}

In this appendix, we carry out a Fourier transform on the configuration space self-energy, in order to isolate the momentum conservation and, thus, identify the expression for the self-energy in momentum space. With this in mind, let us start by performing a Fourier transform on a generic self-energy, $\Sigma(w,v)$, getting
\bea
       \Sigma(k,l)= \int d^4w d^4v e^{i k\cdot w}e^{-i l\cdot v} \Sigma(w,v).
\eea
Since the magnetic field only affects the transverse components and it is expected that the transverse part of the self-energy can be written as the product of the Schwinger phase and a symmetric part  (see {\it e.g.}~\cite{Machet}), we can perform the following decomposition
\bea
       \Sigma(k,l)&=& \int d^4w d^4v e^{i k\cdot w}e^{-i l\cdot v}\Omega(w_\perp,v_\perp) \tilde\Sigma((w-v)_{||},(w-v)_\perp),
\eea
in a similar fashion as for the propagator.
 
Performing the change of variables, $R=w+v$ and $r=w-v$, and working in  the symmetric gauge, where the phase takes the form
\bea
\Omega(w_\perp,v_\perp) = \exp \left(-iq \int_w^v d\xi \cdot  A(\xi)\right) = \exp \left[-i \frac{q B}{4} (-r_1 R_2 + r_2 R_1)\right],
\eea
the parallel part can easily be integrated, getting
\bea
       \Sigma(k,l)= (2 \pi)^2 \delta^{(2)}_{||} (k-l) \int d^2r_\perp d^2R_\perp e^{i R_i[(k-l)_i - \frac{qB}{4}\epsilon_{ij}r_j]} e^{i (k+l)\cdot r_\perp} \tilde\Sigma(k_{||},r_\perp),
\eea
 where the indexes $i, j = 1, 2$. 
 
The integral over $R_i$ is now straightforward and we get
\bea
       \Sigma(k,l)&=& (2 \pi)^2 \delta^{(2)}_{||} (k-l) (2\pi)^2 \left( \frac{4}{qB}\right)^2
       \int d^2r_\perp \delta \left(r_1 +\frac{4}{qB}(k-l)_2\right)  \delta \left(r_2 -\frac{4}{qB}(k-l)_1\right)  e^{i (k+l)\cdot r_\perp} 
       \tilde\Sigma(k_{||},r_\perp)
\nonumber \\
     &=&  (2 \pi)^4 \delta^{(2)}_{||} (k-l) \left( \frac{4}{qB}\right)^2 e^{-\frac{2}{q B}\epsilon_{ij}l_ik_j}
     \int \frac{d^2Q_\perp}{(2\pi)^2}
      e^{i 4[  Q_1(k-l)_2-Q_2(k-l)_1]/qB} 
       \tilde\Sigma\left(k_{||},Q_\perp\right),
\label{app5}
\eea
where, in the last line, we made a Fourier transformation of the self-energy over the transverse coordinates and evaluated the Dirac delta functions.

Reorganizing the above equation in a convenient way, we rewrite it as
\bea
 \nonumber \\
     \Sigma(k,l)&=&   (2 \pi)^4 \delta^{(2)}_{||} (k-l) \frac{1}{qB}\left\{e^{-i\frac{2}{q B}\epsilon_{ij}l_ik_j}
     \int \frac{4d^2Q_\perp}{\pi^2 qB}
      e^{i 4[  Q_1(k-l)_2-Q_2(k-l)_1]/qB} \right\}
       \tilde\Sigma\left(k_{||},Q_\perp\right)
\nonumber \\
     &\equiv&   (2 \pi)^4 \delta^{(2)}_{||} (k-l) \frac{1}{qB}
       \tilde\Sigma\left(k,l_\perp\right),
\label{apendansatz}
\eea
 in such a way, the factor within curly brackets is dimensionless, $ \tilde\Sigma\left(k_{||},Q_\perp\right)$ has the expected dimensions for the self-energy and the remaining factor accounts for the energy-momentum conservation, which in the absence of a magnetic field becomes $(2\pi)^4\delta^{(4)}(k-l)$.
 
\acknowledgements
Support for this work has been received in part from
DGAPA-UNAM under grant numbers PAPIIT-IN117817, PAPIIT-IN118219 and PAPIIT-IN120620. GP thanks the hospitality of Facultad de Ciencias, UNAM, during a sabbatical stay, where this work was completed.


\end{document}